\documentclass[12pt]{iopart}

\usepackage{iopams}
\usepackage{graphicx}%,psfrag}% Include figure files
\begin{document}

\title{Attosecond electronic recollision as field detector}

\author{P.A.~Carpeggiani$^{1,2,3}$, M.~Reduzzi$^{1,2}$, A.~Comby$^{1,4}$, H.~Ahmadi$^{1,5}$, S.~K\"{u}hn$^{6}$, F.~Frassetto$^{7}$, L.~Poletto$^{7}$, D.~Hoff$^{8}$, J.~Ullrich$^{9}$, C.~D.~Schr\"{o}ter$^{10}$, R.~Moshammer$^{10}$, G.~G.~Paulus$^{8}$, G.~Sansone$^{1,2,11}$.}
\address{(1) Dipartimento di Fisica, Politecnico Piazza Leonardo da Vinci 32, 20133 Milano, Italy\\
(2) IFN-CNR, Piazza Leonardo da Vinci 32, 20133 Milano, Italy\\
(3) Institut f\"{u}r Photonik, TU Wien, Gu\ss hausstra\ss e 27-29, 1040 Wien, Austria \\
(4) Universit\'{e} de Bordeaux - CNRS - CEA, CELIA, UMR5107, 351 cours de la Liberation F33405 Talence, France\\
(5) Department of Physical Chemistry, School of Chemistry, College of Science, University of Tehran, Tehran, Iran\\
(6) ELI-ALPS, ELI-Hu Kft., Dugonics ter 13, H-6720 Szeged, Hungary \\
(7) Institute of Photonics and Nanotechnologies, CNR via Trasea 7, 35131 Padova, Italy\\
(8) Institut f\"{u}r Optik und Quantenelektronik, Friedrich-Schiller-Universit\"{a}t Jena, Max-Wien-Platz 1, 07743 Jena, Germany\\
(9) Physikalisch-Technische Bundesanstalt, Bundesallee 100, 38116 Braunschweig, Germany\\
(10) Max-Planck-Institut f\"{u}r Kernphysik, Saupfercheckweg 1, 69117 Heidelberg, Germany\\
(11) Physikalisches Institut, Albert-Ludwigs-Universit\"{a}t Freiburg, Stefan-Meier Strasse 19, 79104 Freiburg, Germany\\
}

\begin{abstract}
We demonstrate the complete reconstruction of the electric field of visible-infrared pulses with energy as low as a few tens of nanojoules. The technique allows for the reconstruction of the instantaneous electric field vector direction and magnitude, thus giving access to the characterisation of pulses with an arbitrary time-dependent polarisation state. The technique combines extreme ultraviolet interferometry with the generation of isolated attosecond pulses.
\end{abstract}

%\maketitle

\section{Introduction}
%\cite{RMP-Krausz-2009, NATPHYS-Corkum-2007, PE-Nisoli-2009}.

%\begin{figure}[htb]
%\centering\includegraphics[width=14cm]{Fig1.eps}
%\caption{}
%\label{Fig1}
%\end{figure}
In the 1980s, the advent of laser pulses with intensities exceeding $10^{14}$~W/cm$^2$ opened up new avenues in strong field laser physics. Phenomena such as Above Threshold Ionization (ATI)~\cite{agostini1979free} and High order Harmonic Generation (HHG)~\cite{ferray1988multiple} became observable. In ATI, an atom absorbs several photons from a low frequency field (low with respect to its ionization potential), releasing a high energy electron. In HHG, instead, the outcome of the laser-atom interaction is the emission of electromagnetic radiation at multiple odd frequencies of the driver.\\
The common ground underlying ATI and HHG was finally understood in the early 1990s, with the formulation of the so called ‘three step model’~\cite{krause1992high,schafer1993above,corkum1993plasma}. Briefly, after tunneling quasi-statically from the atomic Coulomb potential lowered by the electric field of the laser (first step), the electron dynamics evolve under the action of the laser field (second step), in first approximation resembling that of a free charged particle (Strong Field Approximation, SFA~\cite{PRA-Lewenstein-1994}). The electron gains kinetic energy in the external field, and, after laser field reversal, is driven back and recollides with the parent ion (third step).\\
The recollision process can have different outcomes: high-energy ATI photoelectrons are the result of an elastic scattering event, while HHG happens after radiative recombination with the ion. Inelastic scattering is also possible: a second electron can be extracted as a result of the impact, a process called Non-Sequential Double Ionization (NSDI)~\cite{walker1994precision}.\\
The key property of recollision physics is the coherence of the process: the electron is driven back to its parent ion, the one it originated from.\\
The extension of this concept to molecules allowed for the development of a number of techniques that make use of the rescattering electron as a probe of the local environment. Laser Induced Electron Diffraction (LIED)~\cite{meckel2008laser}, stemming from ATI, exploits the fact that structural information on the molecule is encoded in the angular distribution of the rescattered electrons.
Recent results~\cite{wolter2016ultrafast} suggest that the technique holds promise to elucidate molecular structural dynamics with sub-femtosecond temporal resolution and sub-angstrom spatial resolution.\\
High order Harmonic Spectroscopy (HHS), instead, exploits the HHG emission from the molecule as observable. After multiple advances in the last two decades, nowadays the technique allows for refined experiments, such as the measurement and laser control of attosecond charge migration~\cite{kraus2015measurement}.\\
Most importantly, the understanding of recollision physics provided by the ‘three step model’ enabled the birth of attosecond science~\cite{RMP-Krausz-2009}. The radiation emitted via HHG consists of a series of light bursts, each one typically lasting tens to hundreds of attoseconds, in the extreme ultraviolet (XUV) region of the spectrum.
Various techniques have been demonstrated to gate the HHG process~\cite{kienberger2004atomic,sansone2006isolated,feng2009generation}, enabling the generation of a single attosecond pulse, and attosecond metrology~\cite{RMP-Krausz-2009} is now a well established branch of ultrafast optics.\\
Today, these tools enable a number of spectroscopic techniques allowing for the study of ultrafast processes in atoms~\cite{ossiander2017attosecond}, molecules~\cite{calegari2014ultrafast} and solids~\cite{moulet2017soft} with unprecedented temporal resolution.
Recently, the extension of attosecond pulses from the XUV to the soft X-ray region has been demonstrated~\cite{li201753,gaumnitz2017streaking,cousin2017attosecond}. These sources are able to probe atomic core to valence transitions, providing attosecond science with unique chemical selectivity~\cite{pertot2017time,attar2017femtosecond}.

\section{Complete characterization of visible pulses}

Attosecond technology has also led to important advances in the metrology of visible and near-infrared pulses. In general, pulse characterization can be accomplished in the time domain by scanning an optical gate or equivalently in the spectral domain by obtaining the amplitude and phase of all spectral components of the pulse~\cite{Hyyti2016}. The most common implementations in the former category are electro-optical sampling (EOS~\cite{Keiber2016}) or autocorrelation (AC), and, in the latter category, spectral phase interferometry for direct electric field reconstruction (SPIDER), multiphoton intrapulse interference phase scan (MIIPS~\cite{Lozovoy2004}) or dispersion scan (D-Scan~\cite{Miranda2012}). A popular method that operates in both domains is frequency resolved optical gating (FROG~\cite{Trebino2012}). The possibility to generate pulses with a complex structure, i.e. with time-dependent amplitude, frequency and polarization, with extreme bandwidths has fueled a manifold of specialized developments for their characterization. By now, there are schemes available to characterize sub-cycle pulses that exceed octave bandwidths~\cite{Fan2016} and cover a broad range of the visible and infrared spectrum.

Applications of ultrashort pulses in strong field physics require not only the knowledge of field envelope and relative phase (or instantaneous frequency), but also of the carrier-envelope phase (CEP, or absolute phase) of the optical waveform. The absolute phase is also essential to recover the polarization state of the waveform in projection measurements. Schemes to provide this additional information have been implemented through advanced versions of FROG \cite{Snedden2015} or with the \textit{addition} of other techniques such as EOS~\cite{Nomura2013} or stereo-above threshold ionization (Stereo-ATI) \cite{Wittmann2009}. The drawback of these approaches is an enormous increase in system complexity as well as  the number of measurements and retrieval steps implying also limitations on operating bandwidths, pulse energies and reliability. Furthermore, most of these techniques have been demonstrated only for linearly polarized pulses, and while means to obtain also the time-dependent polarization of the field exist, such as tomographic ultrafast retrieval of transverse light E fields (TURTLE)~\cite{Xu2009}, polarization labeled interference versus wavelength of only a glint (POLLIWOG)~\cite{Walecki1997} or vectorial E-field characterization through all-optical, self-referenced (VECTOR)~\cite{Chen2014} and versions of spatially encoded arrangement for temporal analysis by dispersing a pair of light E-fields (SEA TADPOLE)~\cite{Lin2016}, a vigorous effort to develop novel approaches based on sub-cycle non-linear processes such as re-collision and strong-field ionization is ongoing. Similar to EOS, they offer the great advantage of a very direct, unambiguous  and complete measurement of vectorial light wave fields with frequencies reaching up to the ultraviolet.

The currently accepted metrology standard for such pulses is the attosecond streak camera (ASC)~\cite{Mairesse2005, Goulielmakis2004} which is based on the momentum transfer from the test field to photoelectrons during photoionization with a single attosecond light pulse (SAP) which serves as the optical gate. Though the generation of SAPs has meanwhile become routine in several laboratories worldwide it is far from trivial and ASC requires cumbersome and slow electron spectroscopy. Moreover, the fully vectorial field reconstruction additionally calls for angularly resolved electron spectroscopy. Intensities of the test field of $10^{12-14}$~W/cm$^2$ are required in the interaction region, even if FROG algorithms for complete reconstruction of attosecond bursts (FROG-CRAB) are applied in the perturbative regime.

The requirement of compressed attosecond pulses can be circumvented by applying the action of the test pulse directly to the process of HHG at the level of electron wave packet creation and acceleration. Two recently demonstrated methods exploit the exquisite sensitivity of the recollision dipole phase and cutoff energy to sample an unknown field by all-optical means.\\ The first method is the petaherz oscilloscope (PO~\cite{Kim2013a}) which is realized by crossing a pump pulse with the test pulse in the generating medium under a small angle. The test field acting on the electron wave packet during acceleration modifies the recollision dipole phase proportional to its magnitude, however, with a spatially dependent delay across the pump pulse waist. The resultant phase gradient proportional to the temporal derivative of the test field causes a deflection of the emitted XUV beam that is measured in the far field by an angularly resolving spectrometer. The transient of the test pulse is obtained by integrating over the delay between the two pulses and a temporal resolution of $\approx$0.7~fs can be achieved. Higher order terms of the spatial phase distribution due to the wavefront curvature of the XUV beam lead to systematic errors in the simple reconstruction procedure and limit the dynamic range of the method to $<$$10^{12}$~W/cm$^2$. Strategies to overcome this limitation, such as phase retrieval algorithms, have been proposed but not demonstrated to date. Moreover, the measurement of two orthogonal vector components to retrieve the full vector field of the test pulse is not straightforward.\\ The second method is named attosecond resolved interferometric electric-field sampling (ARIES~\cite{Wyatt2016}) and utilizes the modulation of the cutoff photon energy during HHG by the field of the test pulse which is collinearly superimposed on the pump pulse. The temporal resolution arises from the fact that the half-cycle of the pump pulse with the highest field amplitude generates a unique spectral feature that shifts spectrally by an amount proportional to the test field during the electron excursion and amounts to $<$1~fs. Linearity in the mapping is limited to $10^{11-12}$~W/cm$^2$ for typical HHG conditions. Furthermore, the method offers sensitivity to the field vector and is thus capable to reconstruct the full 3D field of a pulse.\\ Finally, recent progress in strong field science of solids has lead to the demonstration of an ASC based on a solid state target medium. In this realization, charge injected into the conduction band of an insulator through a strong pump pulse is detected as an electrical current after acceleration by a perpendicularly polarized, co-propagating test pulse~\cite{Schiffrin2013}. With driving fields of $\approx0.5 \cdot 10^{14}$~W/cm$^2$, test fields of $\approx0.5 \cdot 10^{13}$~W/cm$^2$ could be characterized with a resolution around 1~fs. As with the PO, although the projection of the test field onto a direction defined by the geometry of the solid state device is obtained, measurement of a perpendicular component would require rotation of both the pump pulse and the detector which is feasible but has not been demonstrated.

In a recent work~\cite{Carpeggiani2017}, we introduced a new all-optical technique based on XUV spatial interferometry with isolated attosecond pulses for three-dimensional time domain metrology of ultrashort pulses. The technique is based on the fact that an electron undergoing recollision acts as a directional electric field detector. This allows for the reconstruction of a weak pulse to be probed, when it is introduced as a perturbation of the HHG process. In the present work, we further explore the capabilities of the technique.

%SHG FROG, TURTLE (Xu2009), FROG with CEP (see Nomura2013)
%limitations for material transmission at SHG wavelength
%Spectral interferometry (see Str\"{u}ber2014 thesis), POLLIWOG (see Str\"{u}ber2014 thesis)
%Attosecond streaking, FROG-CRAB (see Kim2017)
%ARIES (see Wyatt2016)
%Petaherz oscilloscope (see Kim2013)
%EOS (see Keiber2016)
%Reviews (see Stibenz2006, Reid2016, Hyyti2014-inLiterature)
%Attosecond Streaking, FROG-CRAB (see Mairesse2005, Goulielmakis2004, 3d: Sabbar2014 diss)

\section{Experiment}
\subsection{Spatial XUV interferometry with isolated attosecond pulses}
\begin{figure}[htbp]
  \centering
  \includegraphics[width=11cm]{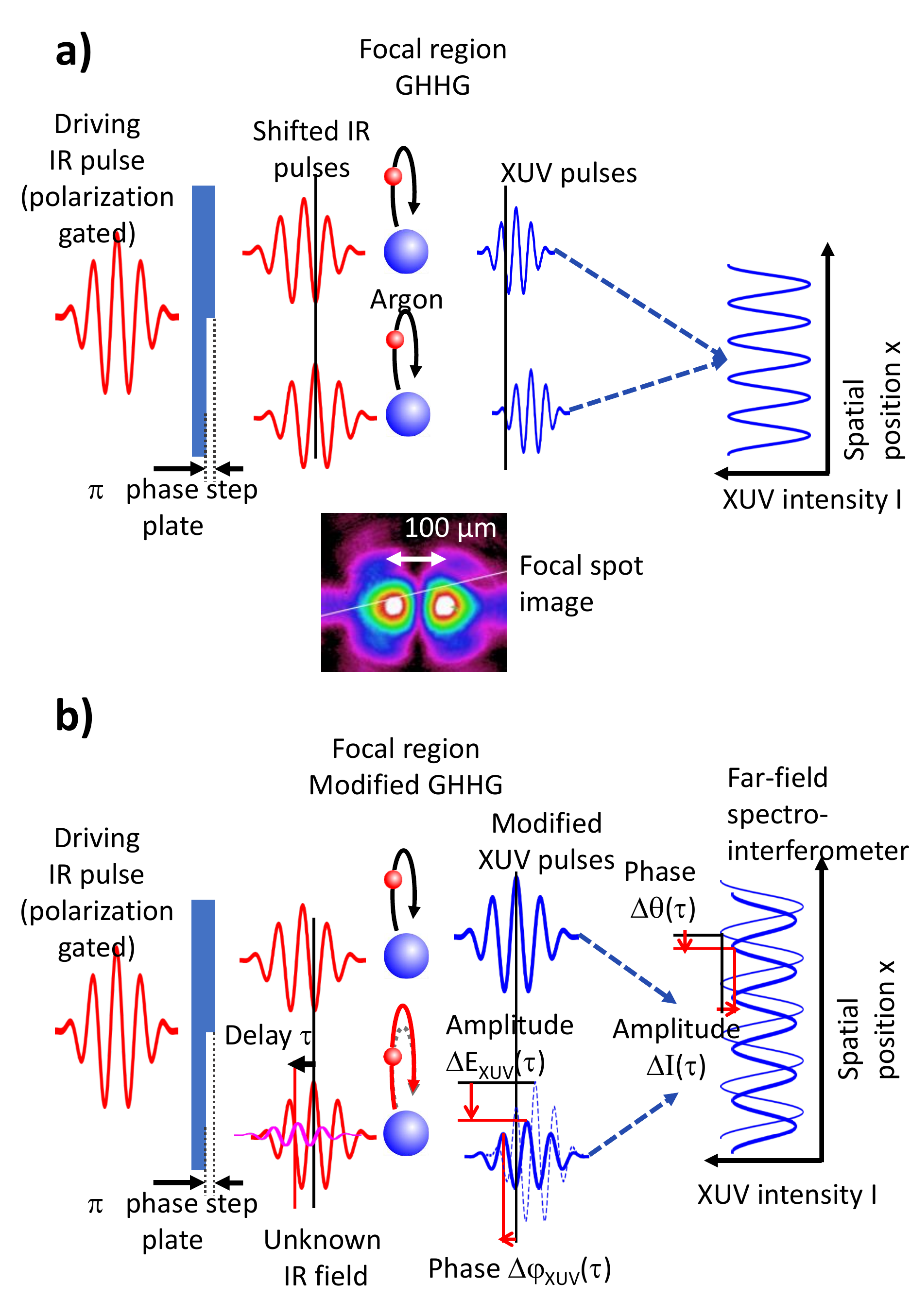}
\caption{(Colors in the online version) a) Principle of XUV spatial interferometry based on isolated attosecond pulses. b) Perturbation introduced by a weak unknown field, which affects the amplitude and the phase of one of the two isolated attosecond pulses.}
\label{Fig1}
\end{figure}
The principle of the technique is shown in Fig.~\ref{Fig1}a,b. Two isolated coherent attosecond pulses are created in two focal spots by two coherent few-cycle pulses with a short time-window of linear polarization. The typical distance between the two spots is on the order of 100~$\mu$m (Fig.~\ref{Fig1}a). The two XUV pulses interfere in the far-field giving rise to a spatial interference pattern, which can be spectrally resolved by means of a XUV spectrometer. The unknown field $E_{unk}(t)$ is overlapped in one of the two focal-spots and its delay $\tau$ with respect to the pair of coherent few-cycle pulses is changed (Fig.~\ref{Fig1}b). The electric field of the unknown pulse modifies the amplitude ($\Delta E_{XUV}(\tau)$) and the phase ($\Delta\varphi_{XUV}(\tau)$) of the attosecond pulse created in the second focal spot, leading to a variation of the contrast ($\Delta I(\tau)$) and the phase ($\Delta\theta(\tau)$) of the fringe interference pattern in the far-field, respectively (Fig.~\ref{Fig1}b). The variation of the contrast $\Delta I(\tau)$ is proportional to the variation of the electric field of the isolated attosecond pulse $\Delta E_{XUV}(\tau)$, while the phase variation $\Delta\varphi(\tau)$ is directly imprinted in the phase of the interference pattern $\Delta\theta(\tau)$.\\
 The ionisation process and the subsequent dynamics of the electronic wave packet leading to the creation of the isolated attosecond pulse is confined to a temporal window of a few hundreds of attosecond. The unknown electric field perturbs only in this time window the harmonic generation process and, therefore, the field can be sampled in time with a sub-femtosecond time resolution.

5-fs, CEP-stable pulses, with an energy of E$\approx$1~mJ at 10~kHz repetition rate were used in the experiment. A small fraction of the IR driving field was used to monitor and stabilize the CEP with a Stereo-ATI setup. We applied the Polarization Gating (PG) technique~\cite{Zhang2004PRA, Sola2006natphys, Zhang2008OE, Sansone2009PRA, Sansone2009bPRA, Hong2010OE} to the driving pulse in order to obtain isolated XUV pulses. The double focus was obtained by a $0-\pi$ diffractive binary element~\cite{PRACamper2014} (see Fig.~\ref{Fig1}). The unknown pulse was generated from the reflection ($\approx4\%$) from a glass plate inserted in the beam path of the linearly polarized few-cycle pulse. The properties of the unknown pulse were modified independently from the main HHG-driving pulse by dedicated elements (iris, quartz plates, variable glass) on its optical path. Its focus is collinearly overlapped with one of the two XUV sources by an off-axis parabola and a drilled mirror. The losses due to the hole in this last mirror reduces the intensity of the unknown pulse in the interaction region.
% whose energy is further reduced first by an iris, and then, for the collinear overlap with the driving pulse, by the reflection on a drilled mirror.
 In the XUV spectrometer used in the experiment, the wavelength and the spatial profile (and therefore the interference fringes) of the XUV radiation were resolved along perpendicular directions. Typical delay scans consisted of 300 steps with integration time 1.2 seconds per step for best data quality, corresponding to 12000 laser shots. Additional information on the experimental setup can be found in Ref.~\cite{Carpeggiani2017}.

\subsection{Characterisation of the CEP of few-cycle pulses}
\begin{figure}[tb]
\centering\includegraphics[width=16cm]{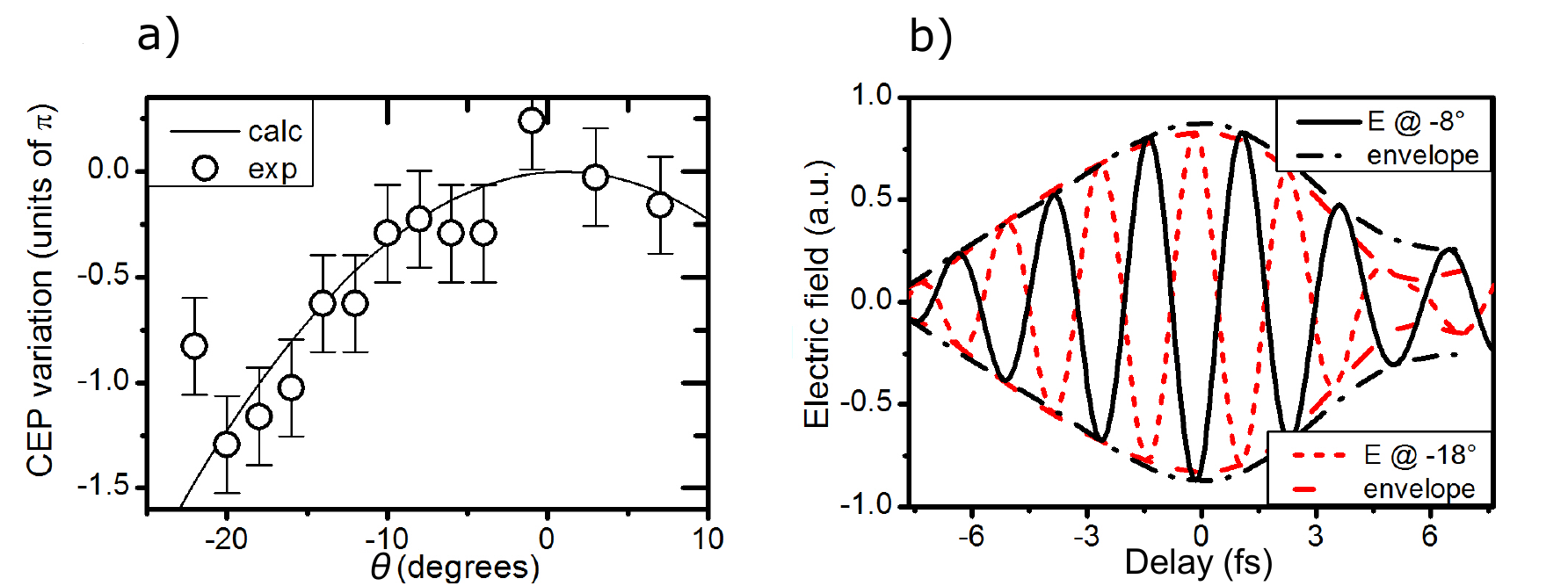}
\caption{
(Colors in the online version). Reconstruction from amplitude modulation of electric fields with different CEP.
 Experimental (circles) and calculated (solid line) CEP variation as a function of the tilting angle $\protect\theta$ of a 1mm thick BK7 glass. The error bars are calculated by the deviations from the calculated value fixing the reduced chi squared to 1 \textbf{(a)}. Electric field and envelope reconstruction for two pulses with a CEP difference of $\protect\pi$. The two electric fields are from the points at $\protect -8^\circ$ (black, solid line for the electric field and dash-dot line for the envelope) and $\protect -18^\circ$ (red, short dash line for the electric field and dash line for the envelope) from the left plot \textbf{(b)}.
}
\label{Fig2}
\end{figure}

In our previous work, we extensively considered the reconstruction of pulses with a complex, time-dependent polarization from the phase modulation~\cite{Carpeggiani2017}. In this work we focus on the experimental results of amplitude-reconstructed pulses.
%The experimental setup is thoroughfully described elsewhere [ref23from intro].
%A reflection of the IR driving field is used to monitor its CEP with a Stereo ATI and to stabilize it with a closed-loop feedback system.
While the CEP of the driving field is actively stabilized, the CEP of the unknown IR field was varied independently by tilting a $1$-mm-thick glass plate. In Fig.~\ref{Fig2}a, the measured variation of CEP is compared with the calculated value according to the equation:
\begin{eqnarray}
\Delta_{CEP}= 2\pi \frac{dn}{d\lambda}(L-d)
\\
1/L=\cos\{\arcsin[\sin\left( \theta \right)/n ]\} / d \nonumber
\end{eqnarray}
where $L=L(d,\theta)$ is the length of the glass crossed by the laser as a function of the tilting angle $\theta$, $d=1$~mm is the thickness of the glass plate, $n=1,5112$ is the index of refraction for BK7 at $\lambda=780$~nm and $dn/d\lambda=-20.887$~mm$^{-1}$ is the corresponding chromatic dispersion. The offset of the experimental values was chosen according to the minimum squares with the calculated curve. The larger deviation of the data points at $-1^\circ$ and $-22^\circ$ is due to random laser fluctuations affecting the quality of those scans.

In Fig.~\ref{Fig2}b we show the reconstructed electric fields, and their relative envelopes, for two pulses with CEP difference close to $\pi$. The two reconstructed fields (black, solid line and red, short-dashed line) correspond to the experimental points of Fig.~\ref{Fig2}a for the values of $\theta=-8^\circ$ and $\theta=-18^\circ$, respectively. These measurements demonstrate that the technique is sensitive to the small field variations induced by a different CEP.

\subsection{Dynamic range of the reconstruction of the field from the amplitude modulation}
\begin{figure}[htb]
\centering\includegraphics[trim=0 0 1cm 0, clip, width=17cm]{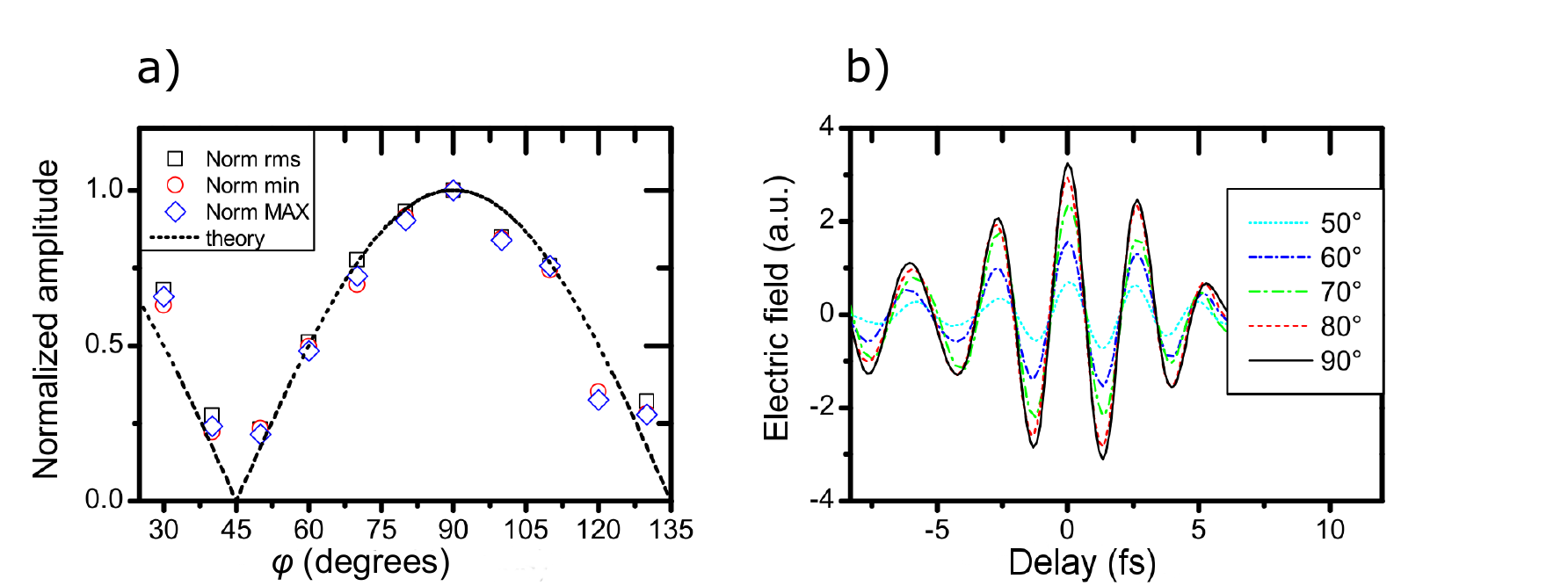}
\caption{
(Colors in the online version). Projection of the electric field on the x-axis as a function of the rotation angle $\protect\phi$ of a $\protect\lambda/2$ waveplate. Root mean square (black squares), positive and negative extreme values (blue diamonds and red circles, respectively) of the electric field, all normalized to 1, and calculated values (black dotted line) \textbf{(a)}. Reconstructed electric fields along the x-axis for five different angles of the $\protect\lambda/2$ waveplate.
$\protect 50^\circ$ (light blue, dot line), $\protect 60^\circ$ (blue, short-dash dot line), $\protect 70^\circ$ (green, dash dot line), $\protect 80^\circ$ (red, short dash line) and $\protect 90^\circ$ (black, solid line) \textbf{(b)}.
}
\label{Fig3}
\end{figure}
As shown in Ref.~\cite{Carpeggiani2017}, this XUV interferometric technique allows only for the reconstruction of the projection of the unknown field along the direction of motion of the electronic wave packet. In order to experimentally demonstrate this property we have used a $\lambda/2$ plate to tune the polarization direction of the unknown pulse, thus changing its projection along the x-axis, which indicates the polarization direction of the attosecond pulse. As shown in Fig.~\ref{Fig3}a, the linearity of the reconstruction is verified by the agreement between the experimental and theoretical amplitude of the electric field. We used three normalized parameters from the reconstructed pulse: the background-subtracted root mean square, which determines the energy of an oscillating signal, and the absolute values of the negative and positive peaks. For the rotation angles of the $\lambda/2$ plate between $50^\circ$ and $90^\circ$ the whole reconstructed fields are shown in Fig.~\ref{Fig3}b.
The cross correlation function measures the similarity of two signals $f(t)$ and $g(t)$ as a function of their displacement $\tau$.
\begin{equation}
(f \star g) = \int_{-\infty}^{+\infty} f^*(t) g(t+\tau) dt
\end{equation}•
For two normalized, identical signals, the maximum value of the function is 1 at $\tau=0$.
%The similarity between two reconstructions is measured by the value of their cross-correlation coefficient. The coefficients for the represented (and normalized) fields are reported in table~\ref{table1}.

%that the amplitude reconstruction can operate in a linear regime, we need to reduce the maximum intensity of the unknown pulse below $10^{12}$~W/cm$^2$ (see sec.\ref{sec:theory}) and to tune the unknown field down to zero.
%IF WE CANNOT GIVE EXACT VALUES, CAN WE SAY THAT WE SWITCH FROM 10\% BS TO 4?
%In order to do so, we take advantage of the directional sensitivity of the technique.

\begin{table}
  \centering
 \begin{tabular}{ || c || c | c | c | c | c ||}
\hline
angle & $50^\circ$ & $60^\circ$ & $70^\circ$ & $80^\circ$ & $90^\circ$ \\
\hline\hline
$50^\circ$ & 1 & 0.98 & 0.99 & 0.99 & 0.91 \\
\hline
$60^\circ$ & 0.98 & 1 & 0.99 & 1.00 & 0.94 \\
\hline
$70^\circ$ & 0.99 & 0.99 & 1 & 0.98 & 0.91 \\
\hline
$80^\circ$ & 0.99 & 1.00 & 0.98 & 1 & 0.94 \\
\hline
$90^\circ$ & 0.91 & 0.94 & 0.91 & 0.94 & 1 \\
\hline
\end{tabular}
\caption{Cross-correlation table of the retrieved electric fields shown in Fig.~\ref{Fig3}}
\label{table1}
\end{table}
In table \ref{table1} the maxima of the cross correlations among the reconstructed signals are reported. Due to the properties of the function, the table is symmetric with respect to the diagonal, and the diagonal elements are equal to~1, as they measure the similarity of a reconstruction with itself (autocorrelation case). The numerical values confirm the high similarity among the reconstructions for different angles of the half-wave plate. With an average value of 0.92, the field reconstructed at $90^\circ$ has the lowest similarity with the others. Even comparing the normalized reconstructions (not shown), this difference is too small to state clearly whether it is due to systematic effects (regime of non-linear response) or just to random variations.
The ratio in amplitude between the smallest and the largest reconstructed electric fields is roughly~1:4, which corresponds to~1:16 in terms of pulse intensity. For the fields shown in Fig.~\ref{Fig3}, the intensities vary from few $10^{9}$~W/cm$^2$ to $\approx10^{11}$~W/cm$^2$.

\section{Theoretical model and simulations}
\label{sec:theory}
The numerical model used to simulate the effect of the unknown field on the high-order harmonic spectra is based on the solution of the Lewenstein model, using the stationary phase method~\cite{PRA-Lewenstein-1994,PRA-Sansone-2004,PRA-Sansone-2009a}.
In the simulations we considered the total electric field ($\mathbf{E}_{tot}$) given by the sum of the field driving the HHG process ($\mathbf{E}_{dr}$) and the unknown field ($\mathbf{E}_{unk}$). Within the stationary phase method, the XUV spectrum $\mathbf{E}(\omega)$ can be expressed as the sum of a finite number of contributions corresponding to the semi-classical paths followed by the photoelectron wave packets leading to the harmonic emission:
\begin{eqnarray}
\mathbf{E}_{XUV}(\omega)&=&\frac{i \
2\pi}{\sqrt{\textrm{det}(S'')}}\Big[\frac{\pi}{\epsilon+i(t_{s}-t'_s)/2}\Big]^{3/2}
 \mathbf{d}^*[\mathbf{p}_s-\mathbf{A}_{tot}(t_s)]\times\nonumber\\ &\times&
\mathbf{E}_{tot}(t'_s)\cdot\mathbf{d}[\mathbf{p}_s-\mathbf{A}_{tot}(t'_s)] \
\exp[-iS(\mathbf{p}_s,t_s,t'_s)+i\omega t_s]
\nonumber\\\label{x(sum)}
\end{eqnarray}
where $S$ is the dipole phase, $\textrm{det}(S'')$ is the determinant of the matrix of the
second time derivative of the phase $(\omega t-S)$ with respect to $t$ and
$t'$ evaluated in correspondence of the saddle points solutions
$(\mathbf{p}_s,t_s,t'_s)$, $\epsilon$ is a regularisation constant, and $\mathbf{d}$ is
the gaussian form of the dipole matrix element~\cite{PRA-Lewenstein-1994}.
The dipole phase $S$ is expressed by:
\begin{equation}
S(p,t,t')=\int_{t'}^{t}\frac{1}{2}\left[\mathbf{p}-\mathbf{A}_{tot}(t^{''})\right]^2dt^{''}+I_p(t-t')
\end{equation}
The saddle points equations are derived by considering the stationary phase points of the the fast oscillating term phase term $(\omega t-S)$:
\begin{eqnarray}
\omega-\frac{[\mathbf{p}_s-\mathbf{A}_{tot}(t_s)]\cdot[\mathbf{p}_s-\mathbf{A}_{tot}(t_s)]}{2}-I_p=0 \nonumber\\
\frac{[\mathbf{p}_s-\mathbf{A}_{tot}(t'_s)]\cdot[\mathbf{p}_s-\mathbf{A}_{tot}(t'_s)]}{2}+I_p=0\label{saddle}
\end{eqnarray}
where $I_p$ is the ionisation potential of the atom, $\omega$ is the harmonic frequency, $\mathbf{A}_{tot}(t)$ is the vector potential associated to $\mathbf{E}_{tot}(t)$, and $t_s$ and $t'_s$ are the stationary values of the recombination and ionisation instants of the electronic wave packet, respectively.
The stationary momentum $\mathbf{p}_s$ is given by:
\begin{equation}
\mathbf{p}_s=\frac{1}{t-t'}\int^{t}_{t'}\mathbf{A}_{tot}(t^{''})dt^{''}\label{ps}
\end{equation}
The complex solutions $(\mathbf{p}_s,t_s,t'_s)$ of Eqs.~(\ref{saddle}-\ref{ps}) identify the most relevant contributions to the XUV spectrum emitted by a single atom.\\
For each temporal delay $\tau$ between the driving and unknown fields, we considered only the solution corresponding to the short path with the highest cut-off frequency~\cite{PRA-Sansone-2004}. The component of the XUV spectrum along the polarisation of the driving field dominates the total signal and we neglect the perpendicular XUV component in the following discussion.\\

\begin{figure}[htb]
\centering\includegraphics[width=15cm]{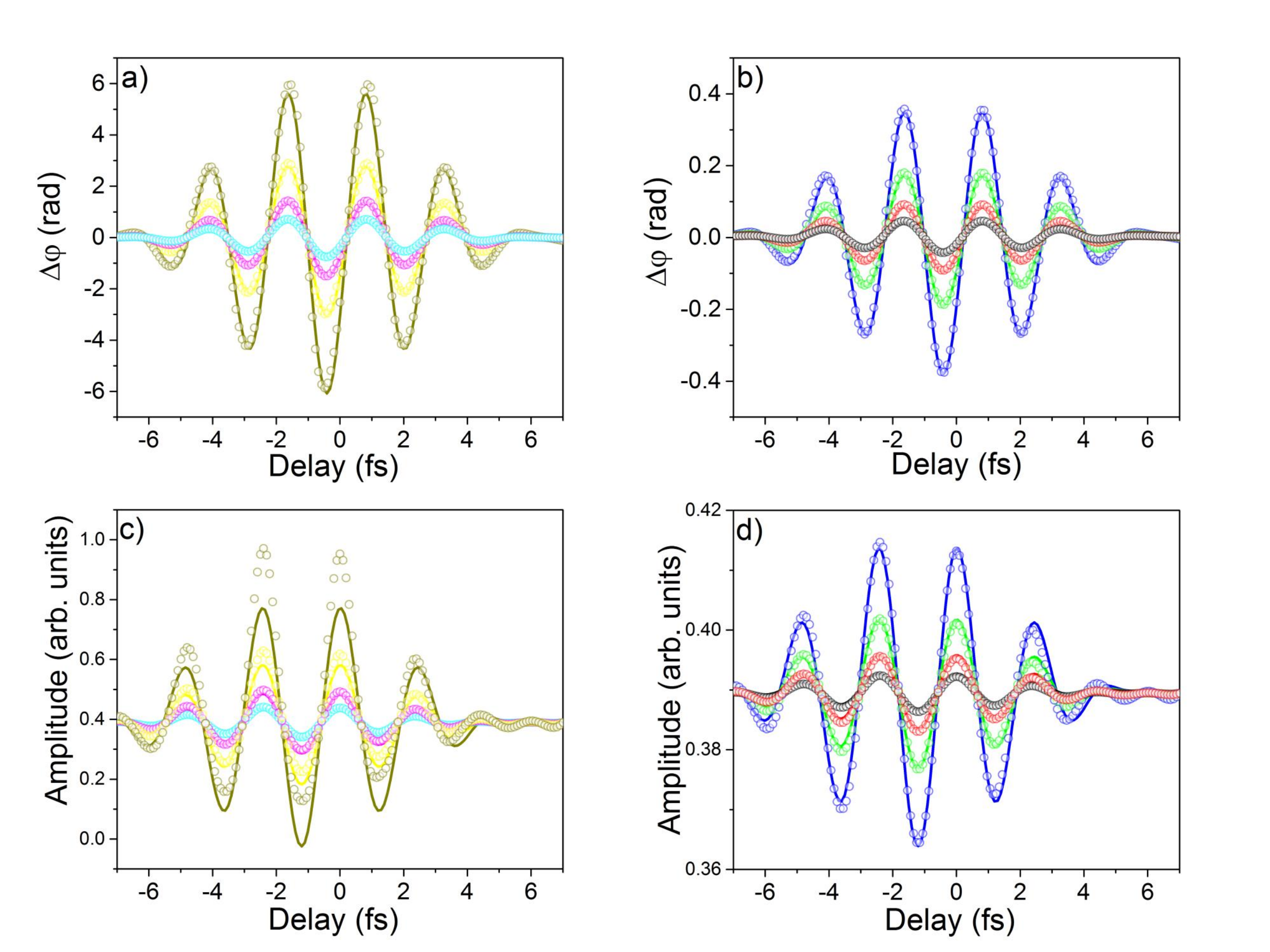}
\caption{(Colors in the online version) Input field (continuous line), phase variation $\Delta\varphi_{XUV}$ (a,b) and amplitude variation $\Delta E_{XUV}$ (c,d) of the harmonics $n=33$ for eight different amplitudes $E_0$ (atomic units) of the unknown field. On the left panels (a,c) the highest values of the electric field: $E_0=$16 (gold), 8 (yellow), 4 (magenta) and $\protect 2\times 10^{-3} a.u.$ (light blue). On the right panels (b,d) the lowest values: $E_0=$1 (blue), 0.5 (green), 0.25 (red) and $\protect 0.12 \times 10^{-3} a.u.$ (black).}
\label{Fig4}
\end{figure}

To validate the possibility to reconstruct the unknown field from the amplitude and phase modulation of the XUV spectra, we have performed simulations for different intensities of the unknown pulse ranging from $I=5.0\times10^8~\mathrm{W/cm^2}$ ($E_0=0.00012~\mathrm{at.~units}$) to $I=9\times10^{12}~\mathrm{W/cm^2}$ ($E_0=0.016~\mathrm{at.~units}$). For each intensity the delay between the unknown and driving field was varied in the range $-8.5$~fs$<\tau<8.5$~fs. We evaluated the effect of the unknown field by considering the amplitude and phase variation of the XUV spectrum $E_{XUV}$. The results are presented in Figs.~\ref{Fig4}a,b and Figs.~\ref{Fig4}c,d for the phase $\Delta\varphi_{XUV}$ and amplitude $\Delta E_{XUV}$, respectively. The unknown electric fields are superimposed in the same graphs (solid line). The input electric field were rescaled according to the peak electric field.
The simulations indicate an excellent agreement between the input electric unknown field and the phase and amplitude modulation of the XUV spectra over a large intensity range. This observation indicates that the perturbation introduces a linear variation (for the intensity ranges considered here) of the characteristics of the XUV spectra, giving access to a direct mapping of the perturbing field. \\
 The reliable reconstruction of the field from the amplitude is restricted to a smaller intensity range between $I=5.0\times10^{8}~\mathrm{W/cm^2}$ and $I=5.6\times10^{11}~\mathrm{W/cm^2}$ ($E_0=0.00012$ and $E_0=0.004~\mathrm{at.~units}$, respectively). For higher intensities, the amplitude modulation deviates from the original field (Fig.~\ref{Fig4}d). This discrepancy could be due to the exponential dependence of the tunneling ionization rate on the instantaneous field, which limits the intensity range where the effect can be considered a perturbation.

We have also verified that pulses with a carrier wavelength as short as $200nm$ can be reproduced with an error between $5\%$ and $15\%$ (evaluated on the single reconstructed electric field value), depending on the duration of the unknown pulse.

The accuracy of the reconstruction will also depend on the intensity of the unknown pulse and, more precisely, on the interplay between its pulse duration, wavelength and intensity. A more exhaustive theoretical analysis varying these three parameters will be needed in order to identify the full parameters space of validity of our approach. \\

\begin{figure}[htb]
\centering\includegraphics[width=10cm]{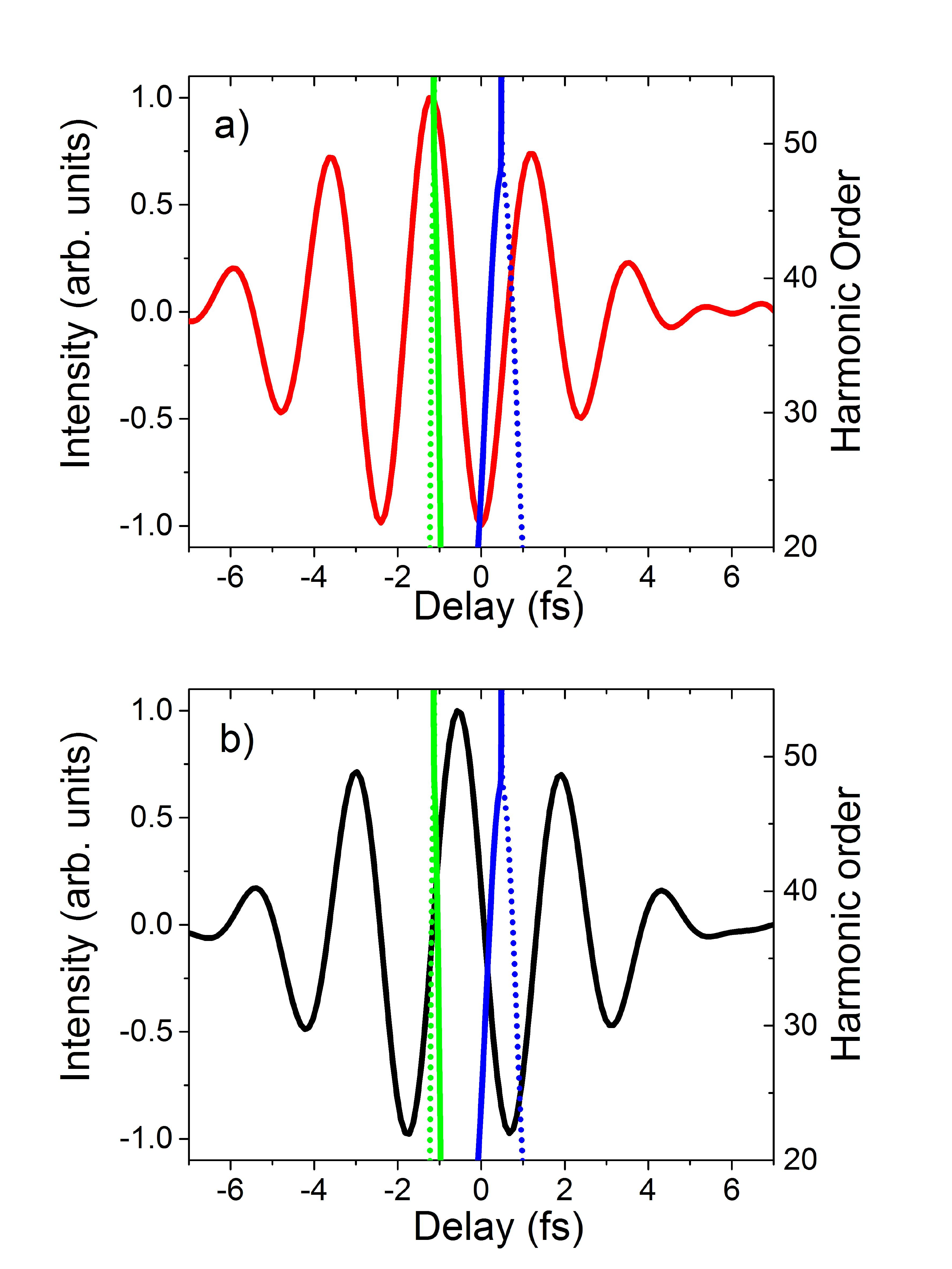}
\caption{(Colors in the online version) Unknown electric field reconstructed from the amplitude (red) (a) and from the phase (black) (b) modulation. Ionization (green) and recombination (blue) curves for the short (solid line) and long (dashes line) paths, leading to the highest harmonic cut-off.}
\label{Fig5}
\end{figure}

It is important to observe that the electric fields reconstructed from the amplitude and phase modulations present a temporal offset, as it can be observed by comparing Fig.~\ref{Fig4}a,c and Fig.~\ref{Fig4}b,d. The same temporal offset was observed experimentally, as shown in Fig.~1e of ref.~\cite{Carpeggiani2017}.
We analyzed the ionization and recombination curves calculated using our model, focusing the attention on those curves leading to the highest cut-off energy. The position in time of these curves is linked only to the oscillation of the driving field.\\
Figure~\ref{Fig5}a shows the unknown field reconstructed from the amplitude modulation (red) together with the ionization (green) and recombination curves (blue) for the central pair of short and long quantum paths. The harmonic order is shown on the right-hand side. We can observe that the maximum of the field reconstructed from the amplitude modulation coincidences with the ionization curves. This observations suggests that the field reconstructed from the amplitude modulation corresponds to the field sampled in time by the ionization process (corresponding to the ionization time of the short path). Therefore, the amplitude modulation can be attributed mainly to the variation of the ionization probability due to the presence of the unknown field, which slightly increases or reduces the (much stronger) driving field.\\
Figure~\ref{Fig5}b shows the field reconstructed from the phase modulation (black line).  We can observe that the maximum of the reconstructed field is placed between the ionization and recombination curves. This suggests that the electric field reconstructed from the phase modulation corresponds to the electric field sampled between the ionization and recombination instants, or, equivalently, it can be considered as a “mean” field between the ionization and recombination instants. This is consistent with the fact that the phase accumulated by the electron wave packet depends on the complete electric field (driving +unknown) during its trajectory in the continuum.\\
Because the amplitude and phase of the XUV interference pattern sample in time the unknown field at different delays, we observe a shift in the field reconstructed from the amplitude and phase modulation.

\section{Conclusions and outlook}
XUV spatial interferometry with isolated attosecond pulses allows for the complete temporal characterisation of visible and near-infrared femtosecond pulses with energies as low as a few tens of nJ. As such this technique could be used for the complete temporal characterisation of the probe pulses typically used in visible pump-probe spectroscopy.
%The complete characterization of the latter pulse would open the possibility for the reconstruction of the complete dielectric wave function of the sample under investigation, in a very similar way of THz time-domain spectroscopy.
Using this approach, it would be possible to measure the small differences imprinted on the electric field of a weak probe pulse interacting with a material with an ongoing electron dynamic. For example, the decay of plasmonic resonance in metallic nanostructure could be investigated using this approach. The complete characterization of the latter pulse would open the possibility for the reconstruction of the complete dielectric wave function of the sample under investigation, in a very similar way of THz time-domain spectroscopy.
\section{Acknowledgments}
This project has received also funding from the European Union's Horizon 2020 research and innovation programme under the Marie Sklodowska-Curie grant agreement no.~641789 MEDEA.

\section*{References}
\bibliography{JPB2018_total}
%\bibliography{JPB_bibliography_mauri}
%\bibitem{PRA-Lewenstein-1994}
%Lewenstein, M., Balcou, P., Ivanov, M. Y., L'Huillier, A. \& Corkum, P. B. Theory of High-Harmonic Generation by Low-Frequency Laser Fields. \textit{Phys. Rev. A} \textbf{49}, 2117-2132 (1994).
%
%\bibitem{PRA-Sansone-2004}
%Sansone, G., Vozzi, C., Stagira, S. \& Nisoli, M. Nonadiabatic quantum path analysis of high-order harmonic generation: Role of the carrier-envelope phase on short and long paths. \textit{Phys. Rev. A} \textbf{70}, 013411 (2004).
%
%\bibitem{PRA-Sansone-2009a}
%Sansone, G. Quantum path analysis of isolated attosecond pulse generation by polarisation gating. \textit{Phys. Rev. A} \textbf{79}, 053410 (2009).
\end{document}